\documentclass[twocolumn,showpacs,preprintnumbers,amsmath,amssymb]{revtex4}
\usepackage{graphicx} % Include figure files
\usepackage{dcolumn}  % Align table columns on decimal point
\usepackage{bm}       % bold math
\begin{document} 
\title{Theory of the waterfall phenomenon in cuprate superconductors}
\author{D. Katagiri$^1$, K. Seki$^1$, R. Eder$^{1,2}$ and Y. Ohta$^1$}
\affiliation{$^1$Department of Phsics, Chiba University, Chiba 263-8522, Japan\\
$^2$Karlsruhe Institut of Technology,
Institut f\"ur Festk\"orperphysik, 76021 Karlsruhe, Germany}
\date{\today}
\begin{abstract}
Based on exact diagonalization and variational cluster approximation
calculations we study the relationship between charge transfer models and
the corresponding single band Hubbard models. 
We present an explanation for the waterfall phenomenon observed in angle
resolved photoemission spectroscopy (ARPES) on cuprate superconductors.
The phenomenon is due to the destructive interference between the phases 
of the O2p orbitals belonging to a given Zhang-Rice singlet
and the Bloch phases of the photohole which occurs in certain regions 
of ${\bf k}$-space. It therefore may be viewed as a direct experimental
visualisation of the Zhang-Rice construction of an effective single
band model for the CuO$_2$ plane.
 \end{abstract} 
%74.72.-h Cuprate superconductors (high-Tc and insulating parent compounds)
%79.60.-i Photoemission and photoelectron spectra
%71.10.Fd Lattice fermion models (Hubbard model, etc.)
\maketitle
\section{Introduction}
The 'waterfall phenomenon' observed in angle resolved photoemission
spectroscopy on cuprate superconductors has attracted some attention.
This phrase summarizes the following  phenomenology
\cite{Ronning2005,Grafetal07,Xieetal,Vallaetal,Panetal,Changetal07,Kordyuketal07,Meevasanaetal,Meevasana,Inosovetal,Ikeda,Moritz,Dong}:
for photon energies around $20\;eV$, where 
photoholes are created predominantly in O2p states\cite{Eastman},
a 'quasiparticle band' can be seen dispersing away from
the Fermi energy as $\Gamma$ is approached 
but then - along the $(1,1)$ direction roughly at 
$(\frac{\pi}{4},\frac{\pi}{4})$ - rapidly looses intensity and cannot be
resolved anymore. Instead there is a 'band' of weak intensity which
seems to drop almost vertically in ${\bf k}$ space
towards higher binding energy, thus creating the
impression of a kink in the quasiparticle band.
Up to binding energies of $\approx 1.0\;eV$ there is a 'black region'
around the $\Gamma$-point with no spectral weight at all.
Finally, at binding energies of around $1.0\;eV$ the 'waterfall' 
seems to merge with one or several bands of high intensity, which often
correspond very well to bands predicted by LDA calculations.\\
In experiments with photon energies 
$\approx 100 eV$ (where an appreciable fraction of the photoholes
is created in Cu3d states\cite{Eastman}) and when the spectra are taken in
higher Brillouin zones\cite{Kordyuketal07,Inosovetal}, however, 
the low energy quasiparticle band
can be resolved all the way to $\Gamma$ and shows no indication
of a kink. This band therefore undoubtedly exists and has
no kink so that the only possible explanation for the 'black region' and the
apparent kink seen at low photon energies are matrix element effects.
This conclusion has in fact been reached by 
Inosov {\em et al.} on the basis of their experimental 
data\cite{Kordyuketal07,Inosovetal}. A very similar conclusion
was also reached by Zhang {\em et al.}\cite{Dong} who showed that
the kink in the quasiparticle band appears only in the second derivative of
the momentum distribution curves 
but is absent in the second derivative of the energy distribution
curves which shows a smooth quasiparticle band instead. They
concluded that the kink is not an intrinsic band dispersion.\\
Some support for this point of view comes from the fact
that the waterfall phenomenon is observed over the whole doping
range, from the antiferromagnetic insulator
to the Fermi-liquid-like
overdoped compounds, so that it is obviously unrelated to any special
features of the electronic structure. Moreover,
the spectrum of a hole in an antiferromagnetic insulator is one of the 
very few reasonably well-understood problems in connection with cuprate 
superconductors. For this special problem good agreement between 
experiment\cite{Wells,Ronning}, 
various approximate calculations\cite{af2,af3,af4,af5,af6,af7,af8,af9},
and exact diagonalization of small 
clusters\cite{szc,dago1,LeungGooding,LeungWells}
has been found and no theory for hole motion in
an antiferromagnet predicts a kink in the quasiparticle band.
Here in particular the work of Leung and Gooding\cite{LeungGooding}
should be mentioned who studied the spectral function for
a single hole in a $32$-site cluster by exact diagonalization
and found a well-defined and smooth quasiparticle band
in excellent agreement with the results of the
self-consistent Born approximation\cite{Rink,MartinezHorsch}. 
Moreover, since plasmons
with energies below the charge transfer gap are not expected
in the insulating compound
and since the coupling to spin excitations is obviously described very
well by the self-consistent Born approximation, 
a hypothetic Bosonic mode can be almost certainly ruled
out as an explanation of the waterfall phenomenon
in the undoped compounds and, due to the near-independence of
the phenomenon on the doping level, also for the entire doping range.\\
One type of matrix-element effect which partly explains
the nonobservation of the quasiparticle band at $\Gamma$
has been discussed by Ronning {\em et al.}\cite{Ronning2005}.
At the $\Gamma$-point 
in the first Brillouin zone the photoelectrons are emitted exactly
perpendicular to the CuO$_2$ plane which we take as  the
$(x-y)$-plane of the coordinate system.
In this situation - and if we neglect small deviations from this symmetry 
in the actual crystal structure -
the experimental setup has $C_{4v}$ symmetry.
The expression for the photocurrent\cite{Feibelman}
involves the dipole matrix element 
$\langle f| {\bf A}\cdot {\bf p} |i\rangle$,
where ${\bf A}$ is the vector potential of the incoming light,
${\bf p}$ is the momentum operator, $|i\rangle$ the initial
and $|f\rangle$ the final state. 
The state $|f\rangle$ differs from  $|i\rangle$ by \\
a) the presence of an electron
in the so-called LEED state, which far from the
surface evolves into a plane wave $\propto e^{ikz}$ and hence
transforms according to the identical representation\\
b)  by the presence
of a hole with momentum $(0,0)$ in a Zhang-Rice singlet.
Since the Zhang-Rice singlet has a Cu3d$_{x^2-y^2}$ orbital as its
'nucleus'\cite{ZhangRice} it has the same symmetry. \\
It follows that $|f\rangle$ and $|i\rangle$ have the same parity
under reflection in the $x-y$ plane so that the dipole matrix
element is zero if ${\bf A}$ is in the CuO$_2$ plane.
On the other hand
$|f\rangle$ and $|i\rangle$ aquire a relative minus-sign
under a rotation by $\frac{\pi}{2}$ around the $z$-axis
so that the dipole matrix element is zero as well
if  ${\bf A}$ is perpendicular to the CuO$_2$ plane.
This means that it is impossible to observe a state with the
character of a Zhang-Rice singlet at $\Gamma$ in normal emission.
As Ronning {\em et al.} pointed out, however, 
this argument  cannot explain the
nonobservation of the quasiparticle band at $\Gamma$ in higher
Brillouin zones, where the photoelectrons are no longer
emitted in the direction perpendicular to the surface.\\
Moreover, a similar effect has been observed in ARPES studies
of the compound SrCuO$_2$ which contains CuO chains. With a photon energy
of $22.4\;eV$ and polarization parallel to the chains - which
generates holes in $\sigma$-bonding O2p orbital -
there is no intensity at $k=0$\cite{Kimetal}. If the photon energy
is increased to $100\;eV$, however, the spinon-band around
$k=0$ can indeed be resolved\cite{Koitzsch}. This behaviour is 
similar to the waterfall effect but in this compound
the CuO$_2$ plaquettes are perpendicular to the surface of the crystal.
We conclude that there must be a second mechanism for the 
extinction of
spectral weight around $\Gamma$ in both, the one and two dimensional
systems.\\
An explanation of the waterfall phenomenon has been given
by Basak {\em et al.}\cite{Basak}.
These authors first showed that the waterfall phenomenon cannot be
reproduced by a calculation of ARPES spectra from
LDA band structures and eigenfunctions alone, a procedure which has
otherwise been found to be highly successful in describing
ARPES spectra of cuprate superconductors\cite{BansilLindroos,Lindroos}. 
Instead, these authors obtained
good agreement with experiment by additionally introducing a self-energy
which describes the coupling to a Bosonic mode. The basic mechanism
for the waterfall and the variation of the ARPES spectra with photon
energy then is bilayer splitting which is
enhanced by the coupling to the Bosonic mode, in particular the
vertical part of the waterfall turns out to be the strongly
renormalized 
bonding combination of the two single-layer wave functions.
This model reproduces the strong changes of ARPES spectra
with photon energy $h\nu$ in the range $60\;eV \rightarrow 80\;eV$
as observed by Inosov {\em et al.}\cite{Kordyuketal07,Inosovetal}
quite well. On the other hand, Inosov {\em et al.} gave a
different explanation for this variation, namely the rapid variation
of the Cu3d photoemission intensity at the Cu $3p \rightarrow 3d$
absorption threshold at $75 \;eV$. Since the strong variation
of photoemission spectra at the   $3p \rightarrow 3d$ threshold
is well established for 3d transition metal oxides\cite{oh} this appears
a  plausible explanation.\\
There have been attempts to reproduce a kink in the band structure
within the framework of a single-band Hubbard or 
t-J model\cite{Macridin,Zemlic,MoritzJohn}.
The high-intensity bands observed near $\Gamma$ at binding
energies of $\approx 1.0\;eV$ thereby are identified with
high energy features observed previously in
exact diagonalization\cite{szc,dago1,LeungGooding} or 
Quantum Monte-Carlo\cite{Carsten} studies for such models. \\
In the present manuscript we take the point of view that the
waterfall phenomenon is a pure matrix-element effect, as
pointed out by Inosov {\em et al.}\cite{Kordyuketal07,Inosovetal}
and Zhang {\em et al.}\cite{Dong}.
Thereby the crucial point is the
nature of the low-energy hole states as Zhang-Rice
singlets. Since the phases of the O2p hole 'within' a Zhang-Rice singlet
correspond to momentum $(\pi,\pi)$\cite{ZhangRice} there is perfect destructive
interference with the phases of a p-like photohole
with momentum $(2n\pi,2m\pi)$ with $n$ and $m$ integer.
The quasiparticle band at $\Gamma$ therefore can be observed only at 
photon energies where the cross section for hole creation
in $d$-orbitals is large - because the destructive interference
occurs only for O2p-photoholes -
and in higher Brillouin zones where the argument
by Ronning {\em et al.}\cite{Ronning2005} does not apply.\\
In section II we will discuss exact diagonalization results for a
1-dimensional charge transfer model. In section
III we discuss the spectra of a 2-dimensional charge transfer model
by the variational cluster approximation (VCA). In section
IV we discuss the experimental relevance of the binding energy of
the Zhang-Rice singlet and section V gives summary and conclusions.
%%%%%%%%%%%%%%%%%%%%%%%%%%%%%%%%%%%%%%%%%%
\section{One dimensional model}
\begin{figure}
\includegraphics[width=\columnwidth]{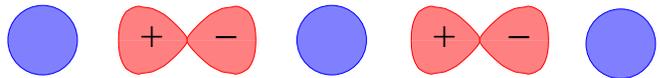}
\caption{\label{fig1}  (Color online)
Possible realization of the model (\ref{ham}).}
\end{figure}

%%%%%%%%%%%%%%%%%%%%%%%%%%%%%%%%%%%%%%%%%%
We study a 1-dimensional (1D) charge transfer model
by exact diagonalization. We choose a 1D model
because we need at least a two-band model and the largest cluster
of a two-band model
we can study in 2D contains $4$ Cu-ions, so that we have virtually
no ${\bf k}$-resolution. As will be seen below, however, a very simple
1D model is sufficient to reproduce the key features of the waterfall
phenomenon. To be more precise, the Hamiltonian reads
\begin{eqnarray}
H &=& t\sum_{i,\sigma}\left[ d_{i,\sigma}^\dagger \;( p_{i+\frac{1}{2},\sigma}^{}
- p_{i-\frac{1}{2},\sigma}^{}) + H.c.\right] \nonumber \\
&& + \Delta \sum_i n_{d,i} + U \sum_i n_{d,i,\uparrow} n_{d,i,\downarrow}.
\label{ham}
\end{eqnarray}
where $d_{i,\sigma}^\dagger$ ($ p_{j,\sigma}^\dagger$) creates a hole
in a $d$-like orbital at site $i$ ($p$-like orbital at site $j$).
We choose $t$ as the unit of energy and use the values
$\Delta=-4$ and $U=8$ throughout. The crucial feature of the
model is that - as in the case for the CuO$_2$ plane-
the $d-p$ hybridization integral has an alternating sign.
A schematic representation of the model is shown in Figure \ref{fig1}.
%%%%%%%%%%%%%%%%%%%%%%%%%%%%%%%%%%%%%%%%%%%%%%%%%%%%%%%%%%%%
%%%%%%%%%%%%%%%%%%%%%%%%%%%%%%%%%%%%%%%%%%%%%%%%%%%%%%%%%%%%
%%%%%%%%%%%%%%%%%%%%%%%%%%%%%%%%%%%%%%%%%%%%%%%%%%%%%%%%%%%%
\begin{figure}
\includegraphics[width=\columnwidth]{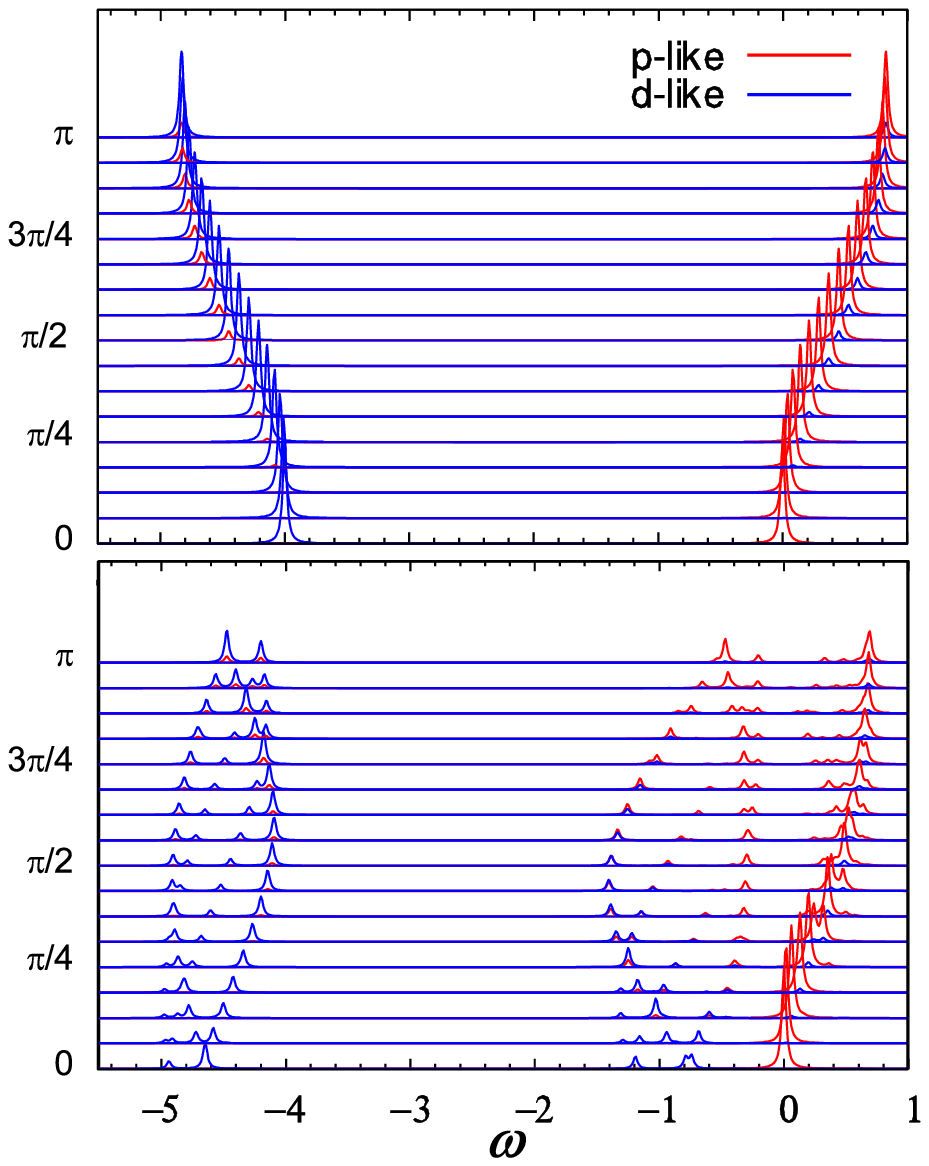}
\caption{\label{fig2}  (Color online)
Single particle spectral function for the model (\ref{ham})
for the half-filled system ($8$ holes in $8$ unit cells). 
The Figure combines
spectra calculated with different boundary conditions to give an impression
of larger systems. The right part of the spectra corresponds to
hole creation (i.e. photoemission) the left part to hole annihilation
(inverse photoemission). The upper figure shows the spectra for $U=0$, 
the lower one  the spectra for $U=8$.}
\end{figure}
%%%%%%%%%%%%%%%%%%%%%%%%%%%%%%%%%%%%%%%%%%%%%%%%%%%%%%%%%%%%
Figure \ref{fig2} shows the single particle spectral function
for the noninteracting case $U=0$ and for the strongly
correlated case $U=8$. These are defined as
\begin{equation}
A_d({\bf k},\omega)= -\frac{1}{\pi}\; \Im \;G_{dd}({\bf k},\omega+i\eta)
\label{specdef}
\end{equation}
where $G_{dd}$ is the $d$-like diagonal element of the $2\times 2$
single-particle Green's function (and analogously for
$A_p({\bf k},\omega)$). The Lorentzian broadening $\eta=0.1$.
In the noninteracting case there are two bands,
one with predominant $p$-character and one with predominant
$d$-character. Since the matrix element of the $p-d$ hybridization
is $\propto \sin(\frac{k}{2})$ there is no $p-d$ hybridization 
for $k=0$ and the states have pure $p$- or $d$-character.
Also, the energy of the $p$-like peak agrees exactly with that of the
$p$-orbital, i.e. $0$. \\
Surprisingly the $p$-like band can be identified 
also in the strongly correlated case, particularly so near $k=0$.
The reason is that the $p$-like Bloch state
with $k=0$ does not mix with the $d$-like Bloch state due to parity 
so that a hole created in this state is unaffected by
the Coulomb repulsion on the $d$-sites.
Accordingly, the energy of this peak still is $0$.
This state is analogous to 
the '1 eV-peaks' in the real cuprate materials\cite{Pothuizenetal}. 
Near  $k=0$ the mixing 
still is small so that $U$ is effectively only a weak perturbation.\\
On the other hand the upper band of predominant $d$-character disappears
completely and is replaced 
by a spectrum which is very similar to that of a 1D single-band
Hubbard model. This can be seen in
Figure \ref{fig3} which shows a close-up of the  low energy region
of the photoemission
(i.e. hole addition) spectrum and compares this to
the photoemisssion spectrum of a single-band Hubbard model.
%%%%%%%%%%%%%%%%%%%%%%%%%%%%%%%%%%%%%%%%%%%%%%%%%%%%%%%%%%%%
\begin{figure}
\includegraphics[width=\columnwidth]{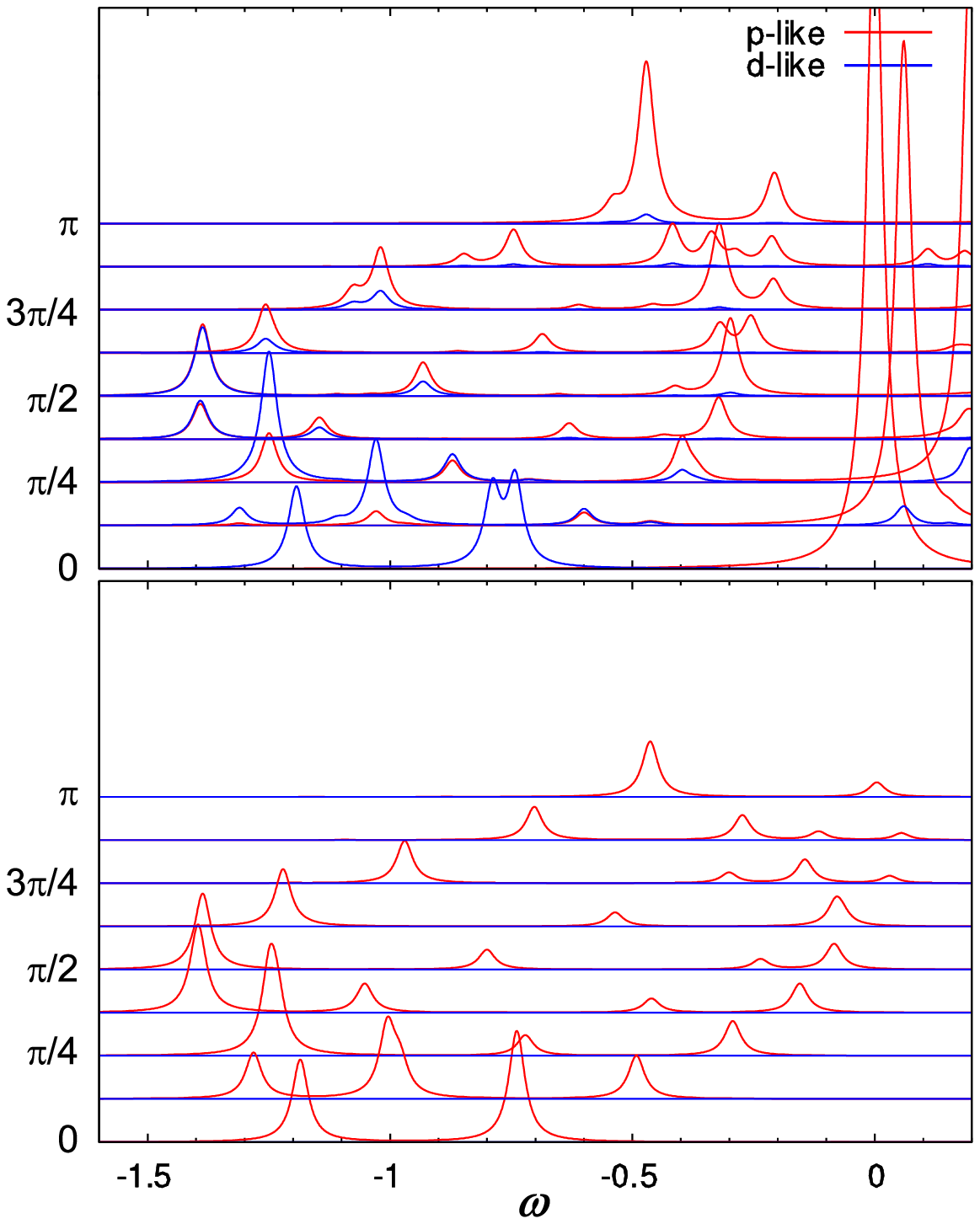}
\caption{\label{fig3}  (Color online)
Top: Closeup of the hole addition part of the spectrum in 
Figure \ref{fig2}. Bottom:
Photoemission spectrum 
for a half-filled single-band Hubbard chain with $8$ sites 
and $U'/t'=10.8$ and $t'=0.335\;t$.
The single-band spectrum has been turned 'upside down' to
be compatible with the hole picture and shifted by $0.87t$.
The figure combines spectra calculated with periodic and antiperiodic 
boundary conditions.}
\end{figure}
%%%%%%%%%%%%%%%%%%%%%%%%%%%%%%%%%%%%%%%%%%%%%%%%%%%%%%%%%%%%
It turns out that for the
above values of $t$, $U$ and $\Delta$ a very good match can
be obtained by choosing the parameters of the single-band Hubbard model
to be $U'/t'=10.8$ and $t'=0.335\;t$.
There is a rather obvious one-to-one correspondence between the
peaks, the different 'holon bands' characteristic for the spectra
of finite clusters of the one-dimensional
Hubbard or t-J model\cite{1dspec} can be identified in both spectra.
The main difference occurs at energies $E$ between $-0.5$ and $0$.
This is most likely the consequence of mixing and level repulsion between 
the single-band Hubbard-like bands and the free-electron-like $p$ band.
Moreover one can see a splitting of some of the peaks in the two-band model.\\
%%%%%%%%%%%%%%%%%%%%%%%%%%%%%%%%%%%%%%%%%%%%%%%%%%%%%%%%%%%%
\begin{figure}
\includegraphics[width=\columnwidth]{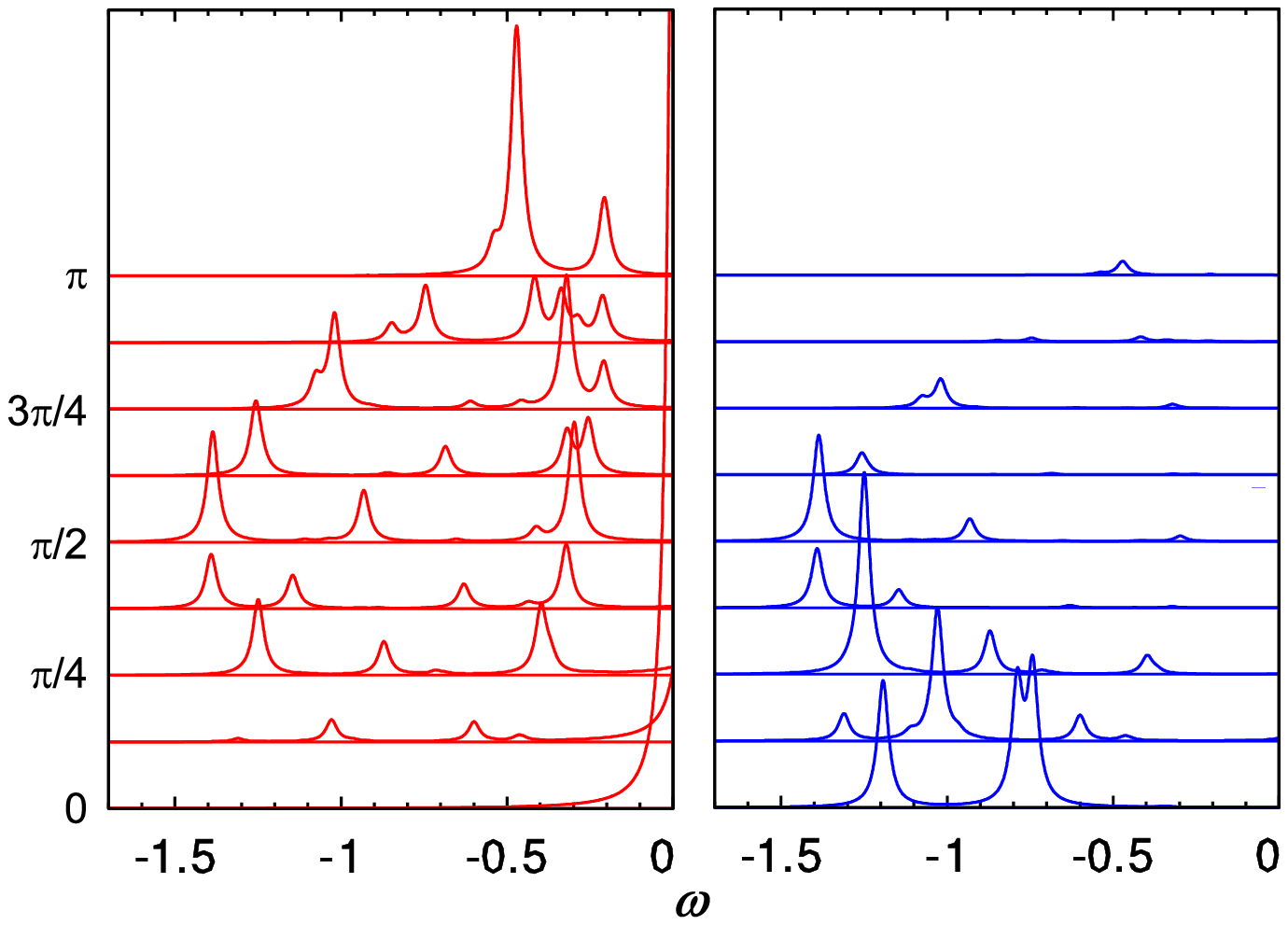}
\caption{\label{fig4}  (Color online)
Closeup of the hole addition part of the spectrum in Figure \ref{fig2}.
The left panel shows the $p$-like spectrum, the right
panel the $d$-like spectrum.}
\end{figure}
%%%%%%%%%%%%%%%%%%%%%%%%%%%%%%%%%%%%%%%%%%%%%%%%%%%%%%%%%%%%
The full interacting spectrum therefore can be modelled 
well by superposing the lower noninteracting $p$ band and a
single band Hubbard spectrum. This indicates that the Zhang-Rice
construction\cite{ZhangRice} for reduction of the low-energy sector
of the two-band model to a single band model also works
well for this 1D model. On the other hand, for higher energies the
2-band model has additional states which correspond to the non-bonding
combination of $p$-orbitals.\\
Figure \ref{fig4} shows the photoemission spectrum
of the two-band model
split into its $p$- and $d$-component. This demonstrates that the
key features\cite{Kordyuketal07,Inosovetal} of the waterfall 
phenomenon can be
seen already in this simple 1D model: Near $\Gamma$ there is
no $p$-like intensity in the single-band Hubbard-like states, 
instead all $p$-like
intensity resides in the free-electron-like band (this is the huge peak at
$E=0$). The $p$-like intensity alone therefore shows the
black region around $k=0$ and
a 'high-energy-kink' whereby the vertical part of the waterfall is due to
the incoherent continuum of the single-particle spectrum of the
single-band Hubbard model. 
In the $d$-like spectrum on the other hand
the quasiparticle band can be followed right up to the $\Gamma$-point.
Already this simple model therefore reproduces the key features of the 
waterfall phenomenon and 
the photon energy dependence of the spectra: the waterfall occurs
in the first Brillouin zone for all photon energies and in higher Brillouin
zones for photon energies where the cross section for
Cu3d is small.\\
To understand the extinction of $p$-like weight
around $\Gamma$ we repeat the Zhang-Rice construction\cite{ZhangRice} 
and consider a single plaquette
- which in the 1D model consists of the $d$-orbital and its two nearest 
neighbor $p$-orbitals -  at the $d$-site
${\bf R}_i$. For one hole, the ground state reads
\begin{eqnarray}
|\Psi_{0,\sigma}^{1}\rangle &=& 
(\;u_1\; p_{i,-,\sigma}^\dagger + v_1\; d_{i\sigma}^\dagger\;)|0\rangle
\nonumber\\
 p_{i,-,\sigma}^\dagger&=& \frac{1}{\sqrt{2}}
( p_{i+\frac{1}{2},\sigma}^\dagger - p_{i-\frac{1}{2},\sigma}^\dagger).
\end{eqnarray}
The coefficients $(u_1,v_1)$ are the GS eigenvector of the matrix
\begin{equation}
H^{1h}=\left( 
\begin{array}{ c c}
0 & \sqrt{2}t\\
 \sqrt{2}t & \Delta
\end{array} \right).
\label{singham}
\end{equation}
The singlet ground state - i.e. the analogue of
a Zhang-Rice singlet in the 1D model - of two holes is
\begin{eqnarray*}
|\Psi_0^{2}\rangle &=& (\;u_2 \;p_{-,\uparrow}^\dagger p_{-,\downarrow}^\dagger
+ \frac{v_2}{\sqrt{2}}\;( d_{\uparrow}^\dagger p_{-,\downarrow}^\dagger
+ p_{-,\uparrow}^\dagger d_{\downarrow}^\dagger\;) 
\nonumber \\
&&\;\;\;\;\;\;+ w_2\; 
d_{\uparrow}^\dagger d_{\downarrow}^\dagger\;) |0\rangle
\end{eqnarray*}
and the coefficients $(u_2,v_2,w_2)$ are the GS eigenvector of the matrix
\begin{equation}
H^{2h}=\left( 
\begin{array}{ c c c}
0 & 2t & 0\\
 2t & \Delta & 2t \\
0& 2t & 2\Delta + U
\end{array} \right).
\label{twoham}
\end{equation}
In the corresponding single-band model, the state 
$|\Psi_{0,\sigma}^{1}\rangle$
corresponds to the site $i$ being occupied
by a spin-$\sigma$ electron, whereas $|\Psi_0^{2}\rangle$
corresponds to a hole at site $i$. The matrix element of the
electron annihilation operator $c_{{\bf k},\uparrow}$
between theses states is $e^{i {\bf k}\cdot{\bf R}_i}$. 
On the other hand, the matrix elements
of the operators $p_{{\bf k},\uparrow}$ and $d_{{\bf k},\uparrow}$
are
\begin{eqnarray}
m_p(k)&=&e^{i {\bf k}\cdot{\bf R}_i}\; \sqrt{2}\;i\sin(\frac{k}{2})\;
(u_1 u_2 + \frac{1}{\sqrt{2}} v_1 v_2) \nonumber \\
m_d(k)&=&e^{i {\bf k}\cdot{\bf R}_i}\; (\frac{1}{\sqrt{2}} u_1 v_2 +
v_1 w_2).
\label{matrixels}
\end{eqnarray}
The crucial term here is the factor of $i\sin(\frac{k}{2})$ which
arises from the overlap of the $p$-like Bloch state with
momentum $k$ and the bonding combination $p_{i,-,\sigma}^\dagger$.
In addition, we have to take into account a shift in
energy of $\epsilon=E_0^{1h}-E_0^{2h}$. This corresponds to 
the binding energy of the ZRS and has no counterpart in the
single-band model.
In simplest terms, we would therefore expect that the
photoemission part of single particle spectral functions
of the two-band model can be obtained from that of the
single band model by
\begin{equation}
A_{p,d}(k,\omega)= |m_{p,d}(k)|^2 A(k,\omega+\epsilon).
\label{approx}
\end{equation}
In this expression several simplifications have been made:
the ZRS is assumed to extend only over one plaquette, which is
probably not correct. This implies that processes
where a ZRS in a plaquette around site $i$ is generated by actually
creating a hole in a neighboring unit cell are neglected.
Moreover the problem of the overlap of Zhang-Rice singlets
in neighboring cells\cite{ZhangRice} is not taken into account either.\\
The energy shift $\epsilon$ which is necessary to match the two spectra
in Figure \ref{fig3}
is found to be $0.87\;t$ - the estimate obtained from the
eigenvalues $E_0^{1h}$ and $E_0^{2h}$ is $1.04\;t$, i.e. reasonably close. 
Figure \ref{fig5} shows the spectra obtained from equation 
(\ref{approx}) compared to the actual
spectra of the two-band model. 
The numerical values of the prefactors are
$u_1 u_2 + \frac{1}{\sqrt{2}} v_1 v_2=0.70$ and
$\frac{1}{\sqrt{2}} u_1 v_2 +v_1 w_2 = 0.50$.
While the agreement is not
really perfect, the qualitative trends are reproduced well,
particularly so near $k=0$ where the hybridization with
the p-like band is weak. 
The main differences occur
for larger $k$-values and may also be due to the fact that
the ZRS is not restricted to one
plaquette and also the hybridization with the $p$-like band
which is absent in the single-band Hubbard model. 
On the other hand, given the simplicity of the procedure for converting 
the single-band spectra into two-band spectra this not so bad and
qualitatively explains the extinction of $p$-like intensity 
around $\Gamma$ at least qualitatively:
this is due to the factor of $\sin(\frac{k}{2})$ in the matrix element
$m_p$, which describes the destructive interference
between the phase factors of the two $p$-orbitals
in the bonding combination $p_{-,\sigma}^\dagger$
and in the electron operator $p_{k,\sigma}^\dagger$. 
This is in turn the consequence of the oscillating sign
of the hopping integral in (\ref{ham}) so that the same mechanism
should also be effective in the 2D CuO$_2$-plane.\\
The exact diagonalization results then can be summarized as follows:
The oscillating sign of the $d-p$ hybridization induces
an oscillating sign also
in the bonding combination of $p$-orbitals around a given
$d$-site. The phase of the $p$-orbitals in the bonding combination
therefore 'locally'
corresponds to a momentum of $\pi$. It follows that 
around $k=2n\pi$ there is destructive interference between the
phases of the bonding combination and the phases of the
photohole, and it is not possible to couple to the
ZRS by hole creation in $p$-orbitals. The ZRS-part
of the spectrum thus becomes extinct in the $p$-like spectrum,
whereas no such extinction occurs in the $d$-like spectrum.
%%%%%%%%%%%%%%%%%%%%%%%%%%%%%%%%%%%%%%%%%%%%%%%%%%%%%%%%%%%%
\begin{figure}
\includegraphics[width=\columnwidth]{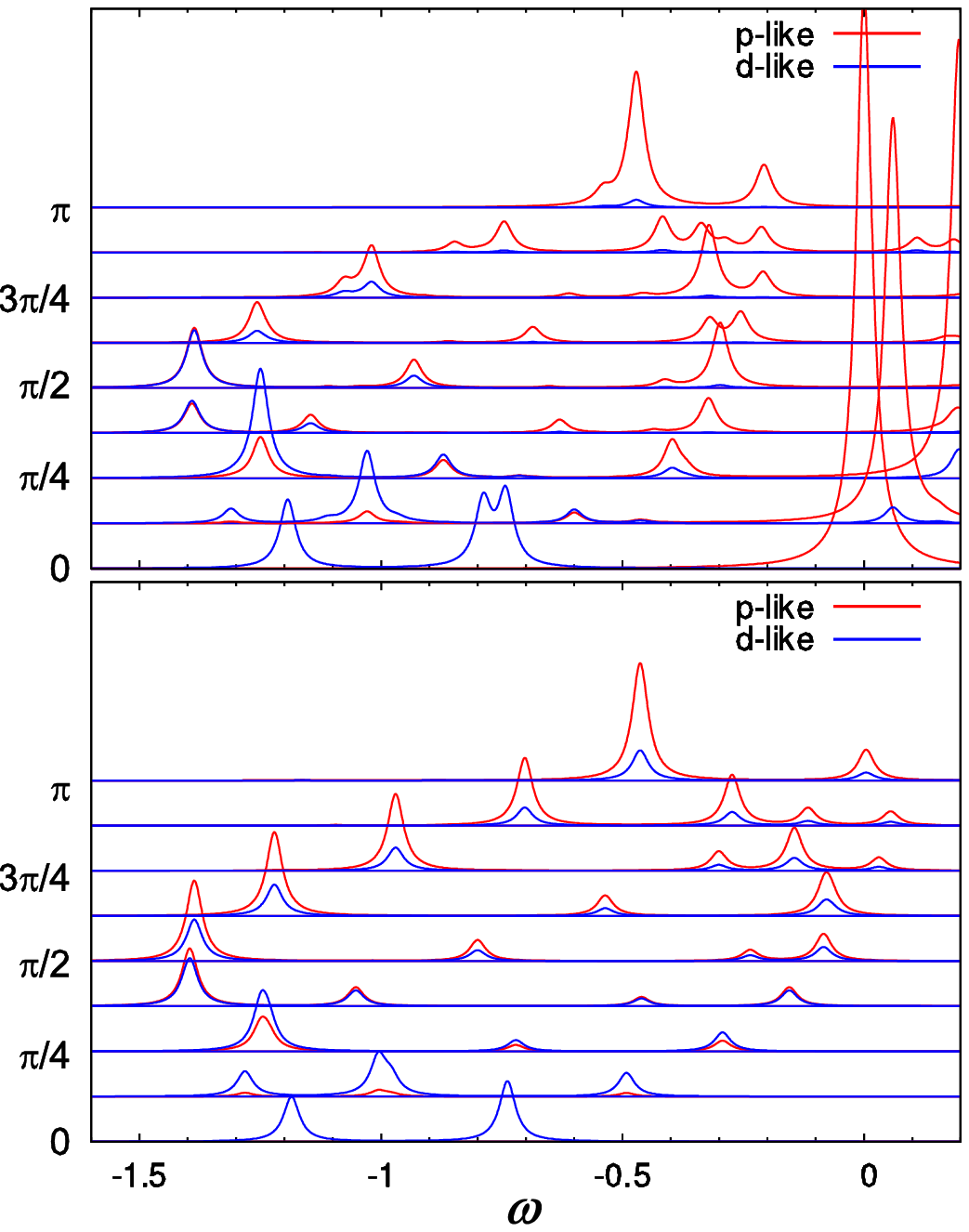}
\caption{\label{fig5}  (Color online)
Closeup of the photoemission spectrum in Figure \ref{fig2} (Top)
compared to the spectra of a single-band Hubbard model corrected 
according to
(\ref{approx}) (Bottom). The energy shift $\epsilon$ is $0.84\;t$
rather than $1.04\;t$ as would be obtained from the single-plaquette
calculation.}
\end{figure}
%%%%%%%%%%%%%%%%%%%%%%%%%%%%%%%%%%%%%%%%%%%%%%%%%%%%%%%%%%%%
To conclude this section we note that
none of the above considerations is limited to half-filling.
Figure \ref{fig6} compares the spectra for the 2-band model
and the single-band model in the hole-doped case,
that means at a hole density of $1.25$. Thereby all
parameters are the same as in Figure \ref{fig4}.
Again, a very good correspondence between the two models exists
and again the extinction of $p$-like weight around
$k=0$ can be clearly seen.\\
%%%%%%%%%%%%%%%%%%%%%%%%%%%%%%%%%%%%%%%%%%%%%%%%%%%%%%%%%%%%
\begin{figure}
\includegraphics[width=\columnwidth]{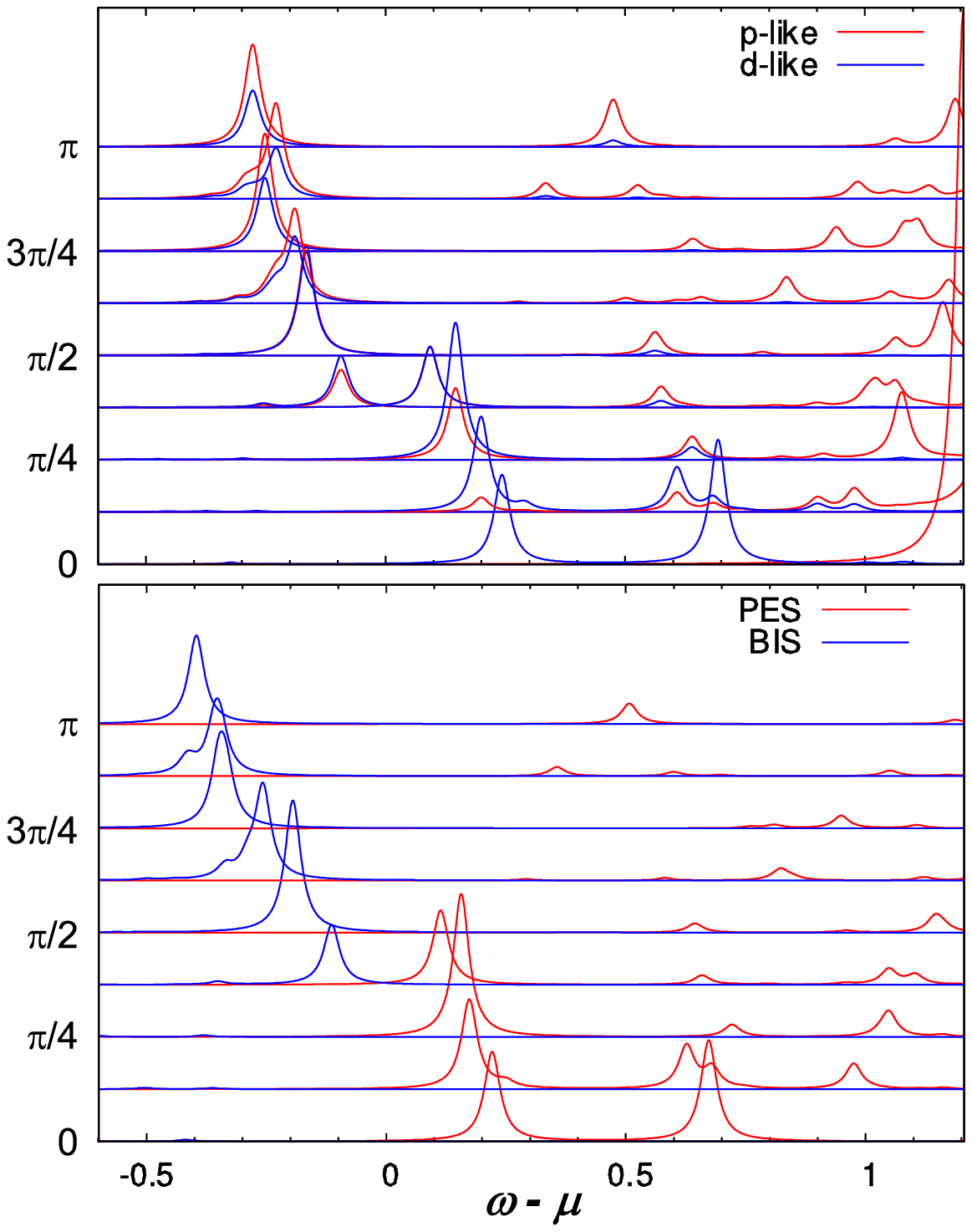}
\caption{\label{fig6}  (Color online)
Low energy photoemission spectrum of the hole-doped
two-band model compared to the spectrum of the corresponding
single-band model. In this figure, the chemical potential is
the zero if energy. }
\end{figure}
%%%%%%%%%%%%%%%%%%%%%%%%%%%%%%%%%%%%%%%%%%%%%%%%%%%%%%%%%%%%
%%%%%%%%%%%%%%%%%%%%%%%%%%%%%%%%%%%%%%%%%%%%
\section{2 Dimensional model}
%%%%%%%%%%%%%%%%%%%%%%%%%%%%%%%%%%%%%%%%%%%%
%%%%%%%%%%%%%%%%%%%%%%%%%%%%%%%%%%%%%%%%%%%%%%%%%%%%%%%%%%%%
\begin{figure}
\includegraphics[width=\columnwidth]{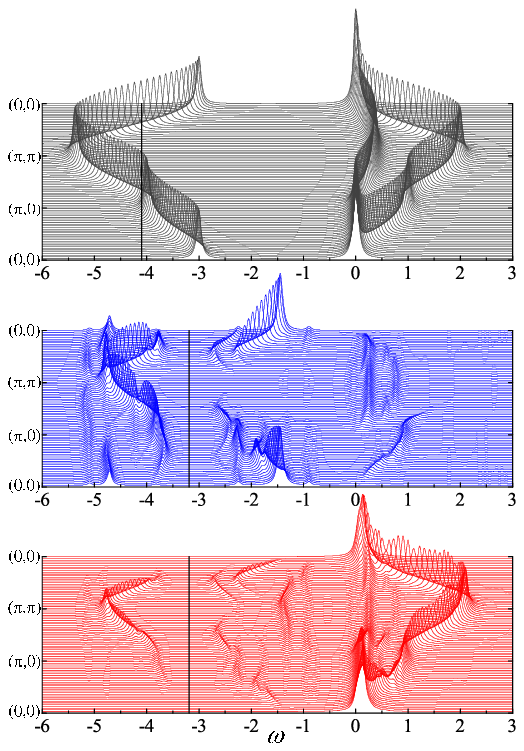}
\caption{\label{fig7}  (Color online)
Top: Sum of $p$-like and $d$-like spectrum for the three-band model 
(\ref{ham2D}) with $U=0$.
Center: $d$-like spectrum for $U=8$
as obtained by the VCA. The spectrum has been multiplied by $2.5$.
Bottom: $p$-like weight for $U=8$.
The spectra are computed for half-filling, that means
$1$ hole/unit cell, the black line denotes the respective
chemical potential.}
\end{figure}
%%%%%%%%%%%%%%%%%%%%%%%%%%%%%%%%%%%%%%%%%%%%%%%%%%%%%%%%%%%%
We now apply the picture gained from the analysis of the one-dimensional 
two-band model to obtain approximate ARPES spectra
for the 2D CuO$_2$-plane. Since it is not possible to
compute the spectra of a 2D three-band model larger
than $2\times 2$ unit cells by exact diagonalization
we switch to the variational cluster
approximation (VCA)\cite{PotthoffI,PotthoffII,Senechal} to compute at least 
approximate spectra.
The VCA uses the fact\cite{LuttingerWard} that the grand canonical potential
of an interacting Fermi system can be expressed as a functional
of the self-energy ${\bf \Sigma}(\omega)$
and is stationary with respect
to variations of ${\bf \Sigma}(\omega)$ at the exact self-energy.
The VCA then uses finite clusters - the so-called 
reference system - to numerically generate
'trial self energies' for an infinite system.
This is described in detail in the 
literature\cite{PotthoffI,PotthoffII,Senechal}
and has turned out to be a very successful method to discuss the 
single-particle spectral functions of correlated electron systems.
We use this method to calculate spectra for the
three-band Hubbard model
\begin{eqnarray}
H &=& 2t_{pd}\sum_{i \in L_d,\sigma} \left(
d_{i,\sigma}^\dagger \;P_{i,\sigma} + H.c \right) \nonumber \\
&&+ 2t_{pp}\sum_{\alpha=x}^y\sum_{j\in L_{\alpha},\sigma}\left(
p_{\alpha,j,\sigma}^{\dagger} Y_{\alpha,j,\sigma} + H.c.\right)\nonumber\\
&& - \Delta \;\sum_{i \in L_{d}} n_{d,i} 
+ U_{dd} \sum_{i \in L_d} n_{d,i,\uparrow} n_{d,i,\downarrow}
\nonumber\\
&& + U_{pp} \sum_{j\in L_{\alpha}} n_{p,j,\uparrow} n_{p,j,\downarrow}.
\label{ham2D}
\end{eqnarray}
$L_d$ denotes the s.c. lattice of Cu3d sites, $L_x$ and $L_y$ the
s.c. lattices of $p_x$ and $p_y$ sites. 
Moreover %Checked! t_pd > 0 with this convention
\begin{equation}
P_{i,\sigma}^\dagger=\frac{1}{2}\left(
p_{x,i-\frac{\hat{x}}{2},\sigma}^\dagger -
p_{x,i+\frac{\hat{x}}{2},\sigma}^\dagger -
p_{y,i-\frac{\hat{y}}{2},\sigma}^\dagger + 
p_{y,i+\frac{\hat{y}}{2},\sigma}^\dagger\right)
\label{bonding}
\end{equation}
is the bonding combination of $p$-orbitals around the
$d$-orbital at site $i$
and %Checked, t_pp > 0
\begin{eqnarray}
Y_{\alpha,i,\sigma}^\dagger&=&\frac{1}{2}(
p_{\bar{\alpha},i+\frac{\hat{x}}{2}-\frac{\hat{y}}{2},\sigma}^\dagger -
p_{\bar{\alpha},i+\frac{\hat{x}}{2}+\frac{\hat{y}}{2},\sigma}^\dagger +
\nonumber \\
&&\;\;\;\;\;\;\;\;\;\;
p_{\bar{\alpha},i-\frac{\hat{x}}{2}+\frac{\hat{y}}{2},\sigma}^\dagger -
p_{\bar{\alpha},i-\frac{\hat{x}}{2}-\frac{\hat{y}}{2},\sigma}^\dagger )
\label{pbonding}
\end{eqnarray}
for $\alpha=x$ denotes
the bonding combination of $p_{y}$ orbitals around a given
$p_{x}$ orbital and vice versa for $\alpha=y$. $\hat{\alpha}$ denotes the 
unit vector in $\alpha$-direction.
The model is again formulated in hole language, i.e.
$d_{i,\sigma}^\dagger$ creates a hole in orbital $i$
and $t_{pd},t_{pp}>0$.
We choose $t_{pd}$ as the unit of energy, the other parameters
are $U_{dd}=8$, $U_{pp}=3$, $\Delta=3$, $t_{pp}=0.5$.
We study this model by the VCA, using
a cluster with $2\times 2$ unit cells (i.e. a square shaped
Cu$_4$O$_8$ cluster) with
4 holes (corresponding to 'half filling') as the reference system
for creating trial self-energies.
Some results obtained in this way for this model
have previously been published by Arrigoni {\em et al.}\cite{Arrigoni}.
Figure \ref{fig7} shows the total spectral weight for the case
case $U=0$ as well as the $p$-like and $d$-like
spectral weight for the interacting case,
$U=8$. The $p$-like spectral weight now is defined as
the sum of the two $p$-like diagonal elements of the
total $3\times 3$ spectral weight matrix.\\
In the noninteracting case, $U=0$, there are three bands.
Switching on the Coulomb repulsion has a very similar effect
as for the 1D model: The two upper (in hole language)
bands with predominant $p$ character remain essentially unchanged, 
whereas the partly filled band of predominant $d$-character
is split into an upper and a lower Hubbard band.
Comparing the $p$ and $d$ spectra in the energy range
$0\rightarrow 2t_{pd}$ the same extinction of 
$p$-like spectral weight around $(0,0)$ can be seen i.e. exactly the
same behaviour as in the 1D model. \\
By analogy with the 1D case we assume that the reason for the extinction
of $p$-like weight again is the matrix element between a plane
wave of $p$ holes and the bonding combination (\ref{bonding}).
We again consider a single-plaquette problem.
For later reference we include a
Coulomb repulsion $U_{pd}$ between $p$ and $d$-holes on
neighboring sites which is set equal to zero for the time being.
The Hamilton matrices for the single and two hole
plaquette problems are:
\begin{equation}
H^{1h}=\left( 
\begin{array}{ c c}
-2t_{pp} & 2t_{pd}\\
 2t_{pd} & \Delta
\end{array} \right)
\label{singham_pl}
\end{equation}
\begin{equation}
H^{2h}=\left( 
\begin{array}{ c c c}
-4t_{pp} +\frac{U_{pp}}{4}& 2\sqrt{2}t_{pd} & 0\\
 2\sqrt{2}t_{pd} & \Delta-2t_{pp} +U_{pd} & 2\sqrt{2}t_{pd} \\
0& 2\sqrt{2}t_{pd} & 2\Delta + U_{dd}
\end{array} \right).
\label{twoham_pl}
\end{equation}
The $p-p$ Coulomb repulsion $\propto U_{pp}$
is treated in mean-field theory.
The ground state eigenvectors of these matrices are again denoted by
$(u_1,v_1)$ and $(u_2,v_2,w_2)$.\\
To discuss the $p$-like photoemission spectrum we need the
matrix element between the bonding combination (\ref{bonding})
and a $p$-like Bloch wave. We write this as
\begin{equation}
m_p=-i \left[ m_x \sin\left(\frac{k_x}{2}\right)
-  m_y \sin\left(\frac{k_y}{2}\right)\right],
\label{bondmat}
\end{equation}
where $m_\alpha$ ($\alpha \in {x,y}$) are matrix elements for
hole creation in a $p_\alpha$ orbital at the origin, which in a real
experiment depend e.g. on the photon polarization and wave vector of
the photoelectrons. In the present calculation $A_p({\bf k},\omega)$
is obtained by calculating the spectra
for creating holes in $p_x$ and $p_y$ orbitals 
separately and adding them - accordingly, the ${\bf k}$-dependent
correction factor in (\ref{approx}) should be replaced by\cite{Wrobel}
\begin{eqnarray}
m_p^2({\bf k})&=& (u_1 u_2 + \frac{1}{\sqrt{2}} v_1 v_2)^2\; 
f({\bf k}) \nonumber \\
f({\bf k}) &=& \sin^2\left(\frac{k_x}{2}\right) + 
\sin^2\left(\frac{k_y}{2}\right)
\label{corrfac}
\end{eqnarray}
To check this we have calclated the photoemission
spectrum of a the single-band Hubbard model
by the VCA using again a $2\times 2$ cluster as reference system.
We use the $2\times 2$ cluster as reference system
in order to make the spectra of the two models as comparable
as possible. By using the same reference
%%%%%%%%%%%%%%%%%%%%%%%%%%%%%%%%%%%%%%%%%%%%%%%%%%%%%%%%%%%%
\begin{figure}
\includegraphics[width=\columnwidth]{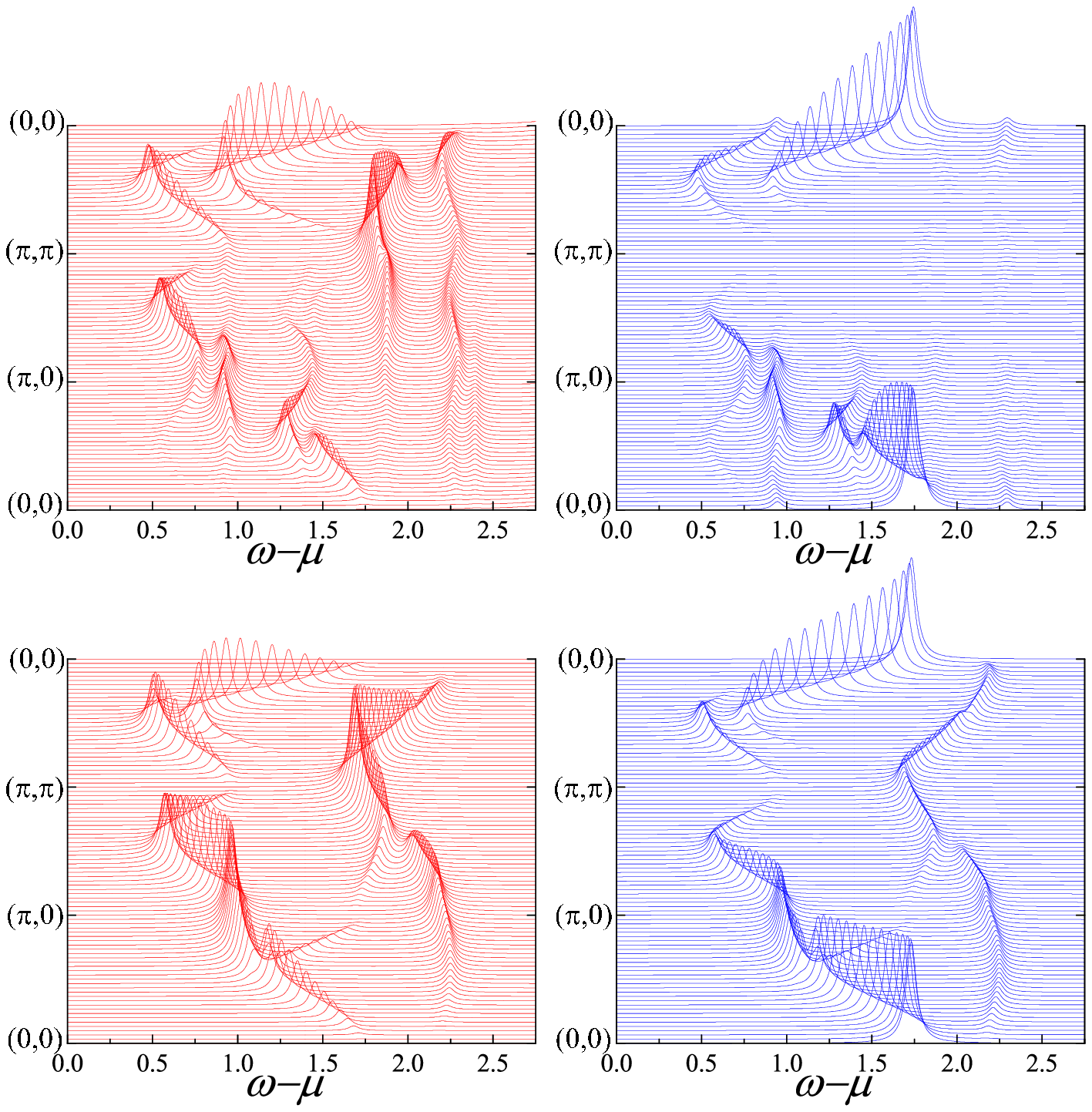}
\caption{\label{fig8}  (Color online)
Right part: $d$-like spectrum of the three-band model
(top) compared to the spectrum of the single band Hubbard model
multiplied by $0.5$ (bottom).\\
Left part: $p$-like spectral weight of the three-band model multiplied
by $2$
(top) compared to the spectrum of the single band Hubbard model
multiplied by $0.5\;f({\bf k})$ in (\ref{corrfac})
(bottom).}
\end{figure}
%%%%%%%%%%%%%%%%%%%%%%%%%%%%%%%%%%%%%%%%%%%%%%%%%%%%%%%%%%%%
system gaps in the VCA bands which originate from the
supercell structure which is inavoidably introduced by the VCA should be
more or less identical in both spectra. The parameter values of
the single band model are again chosen to obtain a good
match with the dispersion of the three-band model.
Thereby in addition to the nearest neighbor hopping integral
$t'$ also hopping integrals $t_1'$ between second-nearest neighbors
and $t_2'$ between third nearest neighbots were included.
A good match was obtained by using $t'/t_{pd}=0.38$,
$t_1'/t'=-0.145$, $t_2'/t'=0.118$ and $U'/t'=10.3$.
Figure \ref{fig8} compares the low energy hole addition
spectrum of the three band model with the (shifted) photoemission
spectrum of the single
band model. The $d$-like spectrum of the three-band model 
should be roughly identical to that of the
single band Hubbard model and the two spectra
on the right part of the figure are indeed quite
similar. There are two main differences:
the weakly dispersive band at $\approx 1.75 t_{pd}$ which can be seen
in the single band model is absent in the spectrum of the three-band model.
This may be a consequence of hybridization with
the two additional $p$-like bands in the three-band model.
Moreover, the spectrum for the three-band model 
has a less smooth dispersion with 
additional gaps along $(0,0)\rightarrow (\pi,0)$. This is likely due
to the lower symmetry of the three-band model:
in the single band Hubbard model, the $2\times 2$ cluster with open
boundary conditions is
equivalent to a $4$-site chain with periodic boundary conditions.
For the two band model a similar reduction is possible, but with
two $p$ orbitals in between any two successive $d$ orbitals.
The resulting  degeneracy of the ligands in the $4$-site chain may give
rise to the additional band splitting.
The left part of the Figure compares the $p$-like spectrum
of the three-band model and the single band model spectrum
corrected by the factor $m_p({\bf k})$. Again, the dispersion of the
spectral weight along comparable bands is very similar in the
two spectra. The energy shift to allign the single-band and
three-band spectra is $\epsilon=2.00\;t_{pd}$. The estimate
obtained from the ground state energies of the matrices
(\ref{singham_pl}) and (\ref{twoham_pl})
is $E_0^{1h}-E_0^{2h} =2.238\;t_{pd}$. The error of
$\approx 10\%$ seems reasonable taking into account the various
approximations made.
To match the intensities of the single-band and three-band spectra
in Figure \ref{fig8} the single band spectra were multiplied by a factor
of $0.5$. The estimates for the correction factors from the
single-plaquette problem are
$(u_1 u_2 + \frac{1}{\sqrt{2}} v_1 v_2)^2=0.56$ and
$(\frac{1}{\sqrt{2}} u_1 v_2 +v_1 w_2)^2 = 0.31$. \\
To summarize the discussion so far: the low-energy
hole addition spectrum of the three-band model can be
obtained to reasonable approximation from an effective single-band model
whereby the $p$-like spectrum needs to be corrected by a simple
${\bf k}$-dependent factor which originates from the interference
between the phases of $p$-orbitals in the bonding combinations
and the phases in a Bloch state with momentum ${\bf k}$.
If this correction is done $p$-like and $d$-like spectra can be
obtained to good approximation
from he spectrum of a single band Hubbard model.\\
We now apply this finding to compute approximate the spectra for
a CuO$_2$ plane, but this time make use of the fact that
in a single-band Hubbard model larger clusters can be 
used as a reference system. We again apply the VCA but thus time we use a
$4\times 4$ cluster with periodic boundary conditions
as reference system. We used a cluster with $2$ holes corresponding
to a hole concentration of $12.5\%$.
It is important to use as large a cluster as possible for the
exact diagonalization step, because
only large clusters reproduce the incoherent continua in the
spectral function sufficiently well and, as will be seen below,
these incoherent continua
are crucial to explain the experimentally observed spectra.
Solution of a $4\times 4$ cluster by exact diagonalization
is only possible with  periodic boundary conditions, which are not
customary in VCA calculations. On the other hand, we do
not want to discuss the phase diagram of the Hubbard model
but we mainly want to obtain the spectral function so 
this is justified. In fact, the spectral function obtained by
the simpler cluster perturbation theory\cite{Senechal1,Senechal2}
is almost exactly the same as the one obtained by VCA.\\
Having obtained an approximate
spectrum for the single-band Hubbard model
we again use (\ref{approx}) together with (\ref{corrfac})
to obtain the $d$-like and $p$-like intensity
for the three-band model. This is shown in Figure \ref{fig9}
for momenta along the $(1,1)$ direction. \\
It can be seen that the Figure reproduces the waterfall effect
quite well. In the $d$-like spectrum the low-energy quasiparticle
band can be seen together with a broad high intensity part
at more negative binding energy. Moreover there is appreciable
incoherent weight around ${\bf k}=(0,0)$. These incoherent continua
are well known from exact 
diagonalization\cite{szc,dago1,LeungGooding,LeungWells}
and self-consistent Born calculations\cite{Rink,MartinezHorsch} 
for the t-J model.
Their intensity decreases with increasing distance from ${\bf k}=(0,0)$.
This decrease is due to the coupling of photoholes to
spin and charge fluctuations\cite{af9}.
In the $p$-like spectrum the  factor $m_p^2({\bf k})$ in (\ref{corrfac})
creates the dark region around ${\bf k}=(0,0)$.
Accordingly, the quasiparticle band seems to
disappear at $\approx (\frac{\pi}{4},\frac{\pi}{4})$. Since the
incoherent weight is reduced by the same factor $m_p^2({\bf k})$, 
it disappears
at the same momentum which creates the impression of a 'band'
which has a kink at $\approx (\frac{\pi}{4},\frac{\pi}{4})$.
At the kink the spectral weight of the band drops sharply.
Qualitatively this is exactly what is seen in the
ARPES spectra which show the waterfall
phenomenon. The chimney-like appearance of the spectum
is due to the fact that the $d$-like spectral weight
can be approximated by a product: $A_{p}(k,\omega)= 
|m_{p}(k)|^2 A(k,\omega+\epsilon)$. For fixed $\omega$
the second factor, $A(k,\omega+\epsilon)$ decreases with $|{\bf k}|$ -
see the left part of
Figure \ref{fig9} - whereas the first factor, $|m_{p}(k)|^2$
vanishes at  ${\bf k}=(0,0)$ and increases with $|{\bf k}|$.
Accordingly, the $p$-like spectral weight must go through a maximum
and this maximum corresponds to the apparent vertical part of the
band. For a quantitative discussion it would be necessary
to take into account also the interference between hole creation
in Cu3d$_{x^2-y^2}$ and O2p orbitals. This however would
necessitate to know the relative magnitude and phase
of the respective dipole matrix elements and this is beyond
the scope of the present paper.
%%%%%%%%%%%%%%%%%%%%%%%%%%%%%%%%%%%%%%%%%%%%%%%%%%%%%%%%%%%%
\begin{figure}
\includegraphics[width=\columnwidth]{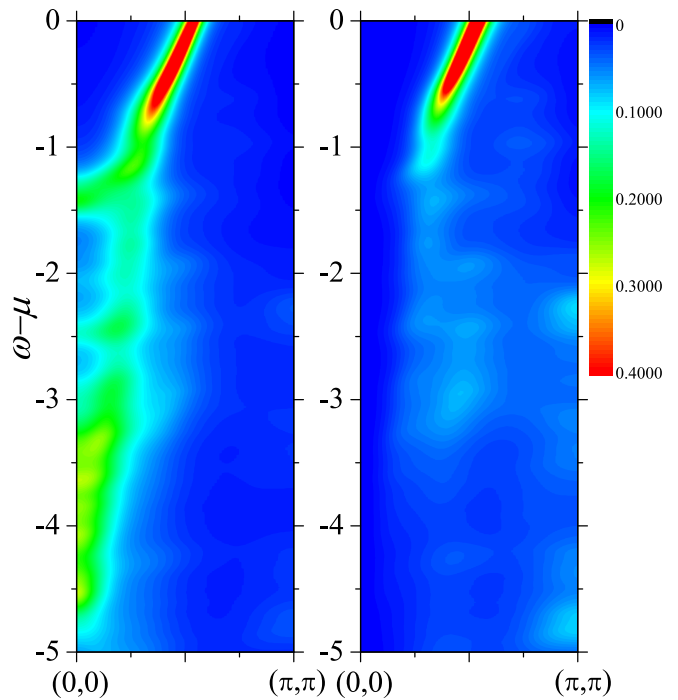}
\caption{\label{fig9}  (Color online)
Left: Spectral function of the the single band Hubbard model
with $U=10$, $t=1$.\\
Right: Spectral function of the the single band Hubbard model
multiplied by $f({\bf k})$ (see (\ref{corrfac})).}
\end{figure}
%%%%%%%%%%%%%%%%%%%%%%%%%%%%%%%%%%%%%%%%%%%%%%%%%%%%%%%%%%%%
%%%%%%%%%%%%%%%%%%%%%%%%%%%%%%%%%%%%%%%%%%%%%%%%%%%%%%%%%%%%
\section{Direct Measurement of the Binding energy of the ZRS}
%%%%%%%%%%%%%%%%%%%%%%%%%%%%%%%%%%%%%%%%%%%%%%%%%%%%%%%%%%%%
Lastly we wish to point out that the energy shift
$\epsilon$ - i.e. the binding energy of the ZRS - does indeed 
have some relevance for the interpretation of experimental
data and has in a sense been observed directly. We refer to the results of
Meevasana {\em et al.}\cite{Meevasanaetal}) who reported an anomalous 
enhancement of the noninteracting bandwidth in Bi2201. These authors
pointed out that the energy difference between an assumed
band bottom at $\Gamma$ and the Fermi energy seems to be
larger than the occupied bandwidth
predicted by LDA calculations and concluded that there is an
anomalous correlation induced band widening rather than
the expected correlation narrowing. \\
We now estimate the position of $\mu$ with respect to
an '1eV' peak as an intrinsic reference energy and show that the Fermi
energy can be obtained quite accurately from
a single-band Hubbard or t-J model by consequent application
of the Zhang-Rice construction\cite{ZhangRice}.
To begin with, Meevasana {\em et al.} observed a band at $\Gamma$
with downward curvature and a binding energy  of $\approx -1\;eV$ 
relative to the Fermi level at $\Gamma$ 
(this is the band labelled B in Figure 1 of Ref. \cite{Meevasanaetal}). 
The authors point out that this is an umklapp of
a band at the $Y$-point, or $(\pi,\pi)$ in a simple cubic 2D model.
This is therefore probably the same state as shown in Fig. 3(c) of
Ref. \cite{Pothuizenetal}, i.e. a state composed of O2p$\pi$ orbitals which
right at $(\pi,\pi)$ has zero hybridization with any of the correlated
Cu3d orbitals. Inspection of Fig. 3(c) of Ref. \cite{Pothuizenetal} shows that
the energy of an {\em electron} in
this state at $(\pi,\pi)$ - and accordingly
its umklapp at $\Gamma$ - is $E_1=\epsilon_p + 4t_{pp}$
where $\epsilon_p$ is the orbital energy for
O2p electrons. Since this is
a single-particle-like state, the corresponding binding
energy in the photoemission spectrum is $E_0^{N}-E_\nu^{N-1}=E_1$.\\
Next, we consider the Fermi energy.
The largest part of the energy shift thereby is the binding energy
of the ZRS, $\epsilon$, which was discussed above. In the matrices
(\ref{singham_pl}) and (\ref{twoham_pl}) the energy of a hole in
an O2p orbital, i.e. $-\epsilon_p$, was chosen as the zero of energy 
so that we have to change $\epsilon \rightarrow \epsilon +\epsilon_p$. 
It remains to add the Fermi energy $\mu_{H}$
of the single-band Hubbard model itself:
\[
\mu=\epsilon_p + \epsilon + \mu_{H}
\]
so that $\mu-E_1=\epsilon + \mu_{H} - 4t_{pp}$
To evaluate $\epsilon$ we use the parameter set given 
by Hybertsen {\em et al.}\cite{Hybertsen} in their Table I.
The only exception is the direct oxygen-oxygen
hopping. Here we use the value $t_{pp}=0.37eV$ which has been
extracted directly from experiment in Ref. \cite{Pothuizenetal}
where it also was found to be consistent with previous estimates.
We then obtain $\epsilon=1.82\;eV$.
The Fermi energy of the single-band Hubbard model
may be estimated from exact diagonalization
results. In a $4\times 4$ cluster this was found to be $1.6\;t$
at $U/t=8$\cite{ortolani} and $1.788\;t$ at $U/t=10$\cite{leung}.
Hybertsen {\em et al.} estimated $U=5.4\;eV$ and $t=0.43\;eV$
so that $U/t=12.6$. We estimate $\mu_{H}\approx 2\;t=0.86\;eV$
so that eventually
$\mu-E_1 = 1.2\;eV$. The value in Figure 1 of Meevasana {\em et al.} is 
$\approx 1\;eV$. The agreement is reasonable given the uncertainty
about some parameters but it is quite obvious
that taking into account
the binding energy $\epsilon$ is indispensable to obtain
a correct estimate of the Fermi energy.\\
The value of $\mu-E_1$ obviously depends on $t_{pp}$ so that small variations
from one compound to the other may well explain the variations observed
by Meevasana {\em at al.}. The Fermi energy of the Hubbard model
will change with doping as well, but these changes are
a faction of $J\approx 120\;meV$. Accordingly, the distance between the
free-electron-like state at $\Gamma$ and the Fermi energy
should always be $\approx 1\;eV$ and this is indeed the case in
Bi2201 and Bi2212\cite{Meevasanaetal}.
All in all one can say that the consequent
application of the ZRS picture can explain the position of the Fermi energy
relative to a 1eV peak - which forms a natural intrinsic reference
energy - quite well.
%%%%%%%%%%%%%%%%%%%%%%%%%%%%%%%%%%%%%%
\section{Summary and discussion}
%%%%%%%%%%%%%%%%%%%%%%%%%%%%%%%%%%%%%%
In summary, we have investigated the relationship between the
single-particle spectra of actual charge-transfer models
and corresponding 'effective' single band Hubbard models.
In the noninteracting case $U=0$ the charge transfer models
have several bands and it was found that those bands with
predominant ligand (i.e. $p$ character in the present models)
are almost unaffected
by the strong correlations. Bands with predominant $d$-character
are split into two Hubbard bands which can be mapped quite well to 
those of an effective single-band Hubbard model.
It turned out that the spectra of $p$ and $d$ electrons to good
approximation can be obtained from that of the single
band Hubbard model by a constant shift in energy - the binding energy
of the Zhang-Rice singlet - and, in the case of the $p$-like spectrum, by
a factor which might be termed the form factor of the
Zhang-Rice singlet.\\
These results give a natural explanation for the waterfall phenomenon
in terms of a pure matrix element effect - as has previously been
inferred by Inosov {\em et al.}\cite{Kordyuketal07,Inosovetal}
and  Zhang {\em et al.}\cite{Dong} from an analysis of their
experimental data.
Here one has to distingish between two different effects:
the first one - pointed out already by 
Ronning {\em et al.}\cite{Ronning2005} - applies only to the special
situation of near-normal emission of photoelectrons. Here
the vanishing of the dipole matrix element
$\langle f|{\bf A}\cdot {\bf p}|i\rangle$
makes it impossible to observe ZRS-derived states.\\
The second matrix element effect is 
the form factor of the Zhang-Rice singlet mentioned above,
which describes the interference
between the phases of the $p$-like photohole and those
of the O2p-orbitals 'within'
a Zhang-Rice singlet - which locally correspond to a state
with momentum $(\pi,\pi)$. This makes it impossible to
couple to a ZRS-derived state by creating a $p$-like photohole
with momenta near $(2n\pi,2m\pi)$ with integer $n$ and $m$.\\
These two simple rules explain 
under which experimental conditions the waterfall is observed
or not: in the hole-doped compounds and
at photon energies where predominantly O2p-like holes are
produced, the quasiparticle band around $\Gamma$ can be
observed neither in the first nor in any higher Brillouin zone
and instead the waterfall appears.
If photon energies where Cu3d holes are generated are used, 
the quasiparticle band cannot be observed in the first BZ, but in higher 
Brillouin zones. In this case the waterfall is absent
and no kink in the quasiparticle band appears.\\
In the electron-doped compounds the quasiparticles
correspond to extra electrons in Cu3d$_{x^2-y^2}$ orbitals so that
the considerations regarding the form factor of the ZRS do not apply.
The argument regarding the vanishing of the dipole matrix
element $\langle f|{\bf A}\cdot {\bf p}|i\rangle$, however, remain
unchanged so that the quasiparticle band  around $\Gamma$
cannot be observed in the first Brillouin zone either - this has
indeed been observed by Ikeda {\em et al.}\cite{Ikeda} and 
Moritz  {\em et al.}\cite{Moritz}. 
It should be possible, however, to observe the full quasiparticle band
without a kink in a higher Brillouin zone.\\
In those cases where the 'dark region' around $\Gamma$ is present, 
the apparent vertical part of the waterfall corresponds to the incoherent 
continua in the
single-particle spectral function of the t-J model. Since these continua
are formed from states which also correspond to Zhang-Rice singlets,
they become extinct near $(2n\pi,2m\pi)$ as well.
Additional evidence comes from the fact that these band portions
show the same dependence on photon polarization as the quasiparticle
band itself\cite{Panetal}.\\
Finally, the high intensity bands observed at $\Gamma$ at binding energies
higher than $\approx 1\;eV$ have no correspondence in a single-band
Hubbard or t-J model - otherwise they would not be observed in normal
emission - but are precisely the bands of predominant
O2p-character which remain unaffected by the strong correlations.
They are analogous to the
$1eV$-peaks observed at high-symmetry points in 
cuprates\cite{Pothuizenetal}
and correspond to $O2p$ drived states which have little or no
hybridization with the strongly correlated $d$-orbitals.
These states therefore are essentially single particle states,
which immediately
explains their much higher intensity as compared to the
quasiparticle band.\\
All in all the present theory indicates that the waterfall phenomenon
constitutes an experimental proof of the Zhang-Rice
construction of a single-band Hubbard or t-J model to describe
the low energy states of the CuO$_2$ planes and in fact
provides a direct visualization of the energy range
in which the states of the real CuO$_2$-plane correspond to 
those of a single-band Hubbard or t-J model.
It shows moreover that the incoherent continua
predicted by various calculations for the t-J or Hubbard model
are indeed observable in experiment in that they
are responsible for the vertical part of the waterfalls themselves.\\
Acknowledgement: K. S. acknowledges the JSPS Research Fellowships
for Young Scientists.
R. E. most gratefully acknowledges the kind hospitality
at the Center for Frontier Science, Chiba University.
This work was supported in part by a Grant-in-Aid for Scientific Research
(Grant No. 22540363) From the Ministry of Education, Culture, Sports,
Science and Technology of Japan. A part of the computations was carried out
at the Research Center for Computational Science, Okazaki Research
Facilities and the Institute for Solid State Physics, University
of Tokyo.


\begin{thebibliography}{}
\bibitem{Ronning2005}
F. Ronning, K. M. Shen, N. P. Armitage, A. Damascelli, D. H. Lu, 
Z. X. Shen, L. L. Miller, and C. Kim, 
Phys. Rev. B 71, 094518 (2005).
\bibitem{Grafetal07}
J. Graf, G.-H. Gweon, K. McElroy, S.Y. Zhou, C. Jozwiak, E. Rotenberg,
A. Bill, T. Sasagawa, H. Eisaki, S. Uchida,  H. Takagi, D.-H. Lee,
and A. Lanzara, Phys. Rev. Lett. {\bf 98}, 067004 (2007).
\bibitem{Xieetal}
B. P. Xie, K. Yang, D. W. Shen, J. F. Zhao, H. W. Ou, J. Wei, S. Y. Gu, 
M. Arita, S. Qiao, H. Namatame, M. Taniguchi, N. Kaneko, H. Eisaki, 
K. D. Tsuei, C. M. Cheng, I. Vobornik, J. Fujii, G. Rossi, Z. Q. Yang, 
and D. L. Feng,
Phys. Rev. Lett. {\bf 98}, 147001 (2007).
\bibitem{Vallaetal}
T. Valla, T. E. Kidd, W.-G. Yin, G. D. Gu, P. D. Johnson, Z.-H. Pan, 
and A. V. Fedorov,
Phys. Rev. Lett.{\bf  98}, 167003 (2007).
\bibitem{Panetal}
Z.-H. Pan, P. Richard, A.V. Fedorov, T. Kondo, T. Takeuchi, S.L. Li, 
Pengcheng Dai, G.D. Gu, W. Ku, Z. Wang, and H. Ding, 
arXiv:cond-mat/0610442 (2006).
\bibitem{Changetal07}
J. Chang, S. Pailhes, M. Shi, M. Manson, T. Claesson, O. Tjernberg, 
J. Voigt, V. Perez, L. Patthey, N. Momono, M. Oda, M. Ido, A. Schnyder, 
C. Mudry, and J. Mesot,
 Phys. Rev. B {\bf 75}, 224508 (2007).
\bibitem{Kordyuketal07}
D. S. Inosov, J. Fink, A. A. Kordyuk, S. V. Borisenko,
V. B. Zabolotnyy, R. Schuster, M. Knupfer, B. Buechner,
R. Follath, H. A. Duerr, W. Eberhardt, V. Hinkov, B. Keimer,
and H. Berger, Phys. Rev. Lett. {\bf 99}, 237002 (2007).
\bibitem{Meevasanaetal}
W. Meevasana, X. J. Zhou, S. Sahrakorpi, W. S. Lee, W. L. Yang, K. Tanaka, 
N. Mannella, T. Yoshida, D. H. Lu, Y. L. Chen, R. H. He, Hsin Lin, S. Komiya, 
Y. Ando, F. Zhou, W. X. Ti, J. W. Xiong, Z. X. Zhao, T. Sasagawa, T. Kakeshita,
K. Fujita, S. Uchida, H. Eisaki, A. Fujimori, Z. Hussain, R. S. Markiewicz, 
A. Bansil, N. Nagaosa, J. Zaanen, T. P. Devereaux, and Z.-X. Shen, 
Phys. Rev. B {\bf 75}, 174506 (2007).
\bibitem{Meevasana}
W. Meevasana, F. Baumberger, K. Tanaka, F. Schmitt, W. R. Dunkel, 
D. H. Lu,  S. K. Mo, H. Eisaki, and Z. X. Shen, 
Phys. Rev. B {\bf 77}, 104506 (2008).
\bibitem{Inosovetal}
D. S. Inosov, R. Schuster, A. A. Kordyuk, J. Fink, S. V. Borisenko, 
V. B. Zabolotnyy, D. V. Evtushinsky, M. Knupfer, B. Büchner, R. Follath, 
and H. Berger,  Phys. Rev. B {\bf 77}, 212504 (2008); 
{\bf 79}, 139901(E) (2009).
\bibitem{Ikeda}
M. Ikeda, T. Yoshida, A. Fujimori, M. Kubota, K. Ono, 
Y. Kaga,  T. Sasagawa, and H. Takagi
Phys. Rev. B {\bf 80}, 184506 (2009).
\bibitem{Moritz}
B. Moritz, F. Schmitt, W. Meevasana, S. Johnston, E. M. Motoyama, M. Greven, 
D. H. Lu, C. Kim, R. T. Scalettar, Z.-X. Shen,
New Journal of Physics {\bf 11}, 093020 (2009).
\bibitem{Dong}
W. Zhang, G. Liu, J. Meng, L. Zhao, H. Liu, X. Dong, W. Lu,
J. S. Wen, Z. J. Xu, G. D. Gu, T. Sasagawa, G. Wang, Y. Zhu,
H. Zhang, Y. Zhou, X. Wang, Z. Zhao, C. Chen, Z. Xu, and
X. J. Zhou, Phys. Rev. Lett. {\bf 101} (2008) 017002.
\bibitem{Eastman}
D. E. Eastman and J. L. Freeouf, Phys. Rev. Lett. {\bf 34}, 395 (1975).
\bibitem{Wells} 
B. O. Wells, Z.-X. Shen, A. Matsuura, D. M. King, M. A. Kastner, M. Greven, 
and R. J. Birgeneau, 
Phys. Rev. Lett.  {\bf 74}, 964 (1995).
\bibitem{Ronning}
F. Ronning, C. Kim, D. L. Feng, D. S. Marshall, A. G. Loeser, L. L. Miller, 
J. N. Eckstein, L. Bozovic, and Z.-X. Shen, Science {\bf 282}, 2067 (1998).
\bibitem{af2}
 B. Kyung and R. A. Ferrell, Phys. Rev. B {\bf 54}, 10125 (1996).
\bibitem{af3}
 F. Lema, and A. A. Aligia, Phys. Rev. B {\bf 55}, 14092 (1997)
\bibitem{af4}
 A. L. Chernyshev, A. V. Dotsenko, and O. P. Sushkov, 
Phys. Rev. B {\bf 49}, 6197 (1994).
\bibitem{af5}
V. I. Belinicher, A. L. Chernyshev, A. V. Dotsenko, and O. P. Sushkov, 
Phys. Rev. B {\bf 51}, 6076 (1995).
\bibitem{af6}
J. Bała, A. M. Oleś, and J. Zaanen, Phys. Rev. B {\bf 52}, 4597 (1995).
\bibitem{af7}
N. M. Plakida, V. S. Oudovenko, P. Horsch, and A. I. Liechtenstein, 
Phys. Rev. B {\bf 55}, R11997 (1997).
\bibitem{af8}
V. I. Belinicher, A. L. Chernyshev, and V. A. Shubin, 
Phys. Rev. B {\bf 56}, 3381 (1997).
\bibitem{af9}
O. P. Sushkov, G. A. Sawatzky, R. Eder, and H. Eskes, 
Phys. Rev. B {\bf 56}, 11769 (1997).
\bibitem{szc}
K. J. von Szczepanski, P. Horsch, W. Stephan and M. Ziegler, 
Phys. Rev. B {\bf 41}, 2017 (1990).
\bibitem{dago1}
E. Dagotto, R. Joynt, A. Moreo, S. Bacci and E. Gagliano, 
Phys. Rev. B {\bf 41}, 9049 (1990).
\bibitem{LeungGooding}
P. W. Leung and R. J. Gooding
Phys. Rev. B {\bf 52}, R15711 (1995).
\bibitem{LeungWells}
P. W. Leung, B. O. Wells, and R. J. Gooding
Phys. Rev. B {\bf 56}, 6320 (1997).
\bibitem{Kimetal}
C. Kim, Z.-X. Shen N. Motoyama, H. Eisaki, S. Uchida, 
T. Tohyama, and S. Maekawa 
Phys. Rev. B {\bf 56}, 15589, (1997).
\bibitem{Koitzsch}
A. Koitzsch, S. V. Borisenko, J. Geck, V. B. Zabolotnyy, M. Knupfer, 
J. Fink, P. Ribeiro, B. Büchner, and R. Follath
Phys. Rev. B {\bf 73}, 201101 (2006).
\bibitem{Basak}
S. Basak, T. Das, H. Lin, J. Nieminen, 
M. Lindroos, R. S. Markiewicz, and A. Bansil,
Phys. Rev. B {\bf 80}, 214520 (2009).
\bibitem{BansilLindroos}
A. Bansil and M. Lindroos, Phys. Rev. Lett. {\bf 83}, 5154 (1999).
\bibitem{Lindroos}
M. Lindroos, S. Sahrakorpi, and A. Bansil, 
Phys. Rev. B {\bf 65}, 054514 (2002).
\bibitem{oh}
S.-J. Oh, J. W. Allen, I. Lindau, and J. C. Mikkelsen, Jr., 
Phys. Rev. B {\bf 26}, 4845 (1982).
\bibitem{Macridin}
A. Macridin, M. Jarrell, T. Maier, and D. J. Scalapino, 
Phys. Rev. Lett. {\bf 99}, 237001 (2007).
\bibitem{Zemlic}
M. M. Zemljic, P. Prelovsek, and T. Tohyama
Phys. Rev. Lett. {\bf 100}, 036402 (2008).
\bibitem{MoritzJohn}
B. Moritz, S. Johnston, and T. P. Deveraux, 
preprint arXiv:1004.4685.
\bibitem{Carsten}
C. Gr\"ober, R. Eder, and W. Hanke, Phys. Rev. B {\bf 62}, 4336 (2000).
\bibitem{Feibelman}
 P.J. Feibelman and D.E. Eastman, Phys. Rev. B {\bf 10}, 4932 (1974).
\bibitem{ZhangRice}
F. C. Zhang and T. M. Rice, Phys. Rev. B {\bf 37}, 3759 (1988).
\bibitem{ShenNiO}
Z. X. Shen {\em et al.}, Phys. Rev. B {\bf 44}, 3604 (1991).
\bibitem{Pothuizenetal}
J. J. Pothuizen, R. Eder, N. T. Hien, M. Matoba, A. A. Menovsky,
and G. A. Sawatzky, Phys. Rev. Lett. {\bf 78}, 717 (1997).
Phys. Rev. B {\bf 41}, 9049 (1990).
\bibitem{1dspec}
R. Eder and Y. Ohta,
Phys. Rev. B {\bf 56}, 2542 (1997).
\bibitem{Rink}
S. Schmitt Rink, C. M. Varma and A. E. Ruckenstein, 
Phys. Rev. Lett. {\bf 60}, 2793 (1988).
\bibitem{MartinezHorsch}
G. Martinez and P. Horsch,
Phys. Rev. B {\bf 44}, 317 (1991)
\bibitem{Hybertsen}
M. S. Hybertsen, E. B. Stechel, M. Schlüter and D. R. Jennison, 
Phys. Rev. B {\bf 41}, 11 068 (1990).
\bibitem{ortolani}
E. Dagotto, F. Ortolani, and D. Scalapino, 
Phys. Rev. B {\bf 46}, 3183 (1992).
\bibitem{leung}
P. W. Leung, Z. Liu, E. Manousakis, M. A. Novotny, and P. E. Oppenheimer,
Phys. Rev. B {\bf 46}, 11779 (1992).
\bibitem{PotthoffI}
M. Potthoff, Eur. Phys. J. B{\bf 36}, 335 (2003)
\bibitem{PotthoffII}
M. Potthoff, Eur. Phys. J. B {\bf 32}, 429 (2003).
\bibitem{LuttingerWard}
J. M. Luttinger and J. C. Ward,
Phys. Rev.{\bf  118}, 1417 (1960).
\bibitem{Senechal}
D. Senechal, arXiv:00806.2690.
\bibitem{Arrigoni}
E. Arrigoni, M. Aichhorn, M. Daghofer, and W. Hanke,
New. J. Phys. {\bf 11}, 055066 (2009).
\bibitem{Wrobel}
P. Wr\'obel, W. Suleja, and R. Eder
Phys. Rev. B {\bf 78}, 064501 (2008).
\bibitem{Senechal1}
D. Senechal, D. Perez, and M. Pioro-Ladriere, 
Phys. Rev. Lett. {\bf 84}, 522 (2000).
\bibitem{Senechal2}
 D. Senechal, D. Perez, and D. Plouffe, 
Phys. Rev. B {\bf 66}, 075129 (2002).
\end{thebibliography}
\end{document}